\documentclass[preprint,prb,amsmath,amssymb,floatfix]{revtex4-1}

\usepackage{amsmath}%
\usepackage{amsfonts}%
\usepackage{amssymb}%
\usepackage{graphicx}
\usepackage{dcolumn}
\usepackage{bm}
\usepackage{mathrsfs}

\begin{document}

\title{Slow reflection and two-photon generation of microcavity exciton-polaritons}

\author{Mark Steger}
\email{mds71@pitt.edu.}
\author{Chitra Gautham}
\author{David W. Snoke}
 
\affiliation{Department of Physics and Astronomy, University of Pittsburgh, 3941 O'Hara Street, Pittsburgh, Pennsylvania 15260, USA}

\author{Loren Pfeiffer}
\author{Ken West}

\affiliation{Department of Electrical Engineering, Princeton University, Princeton, New Jersey 08544, USA}

\date{\today}
						
\begin{abstract}

We resonantly inject polaritons into a microcavity and track them in time and space as they feel a force due to the cavity gradient. This is an example of ``slow reflection,'' as the polaritons, which can be viewed as renormalized photons, slow down to zero velocity and then move back in the opposite direction. These measurements accurately measure the lifetime of the polaritons in our samples, which is 180 $\pm$ 10 ps, corresponding to a cavity leakage time of 135 ps and a cavity $Q$ of 320,000.  Such long-lived polaritons propagate millimeters in these wedge-shaped microcavities.  Additionally, we generate polaritons by two-photon excitation directly into the polariton states, allowing the possibility of modulation of the two-photon absorption by a polariton condensate.

\end{abstract}


\maketitle

\section{Introduction}

Since the initial observation of exciton-polaritons in a strongly coupled microcavity in 1992 \cite{Weisbuch1992}, a wide range of quantum many-body effects have been observed in polariton fluids such as Bose-Einstein Condensation \cite{Kasprzak2006, Balili2007}, and superfluidity exhibiting quantized vortices \cite{Lagoudakis2011} and solitons\cite{Amo2011}.  Most of these results have been interpreted in terms of nonequilibrium Bose gas theory, because the thermalization of the polaritons has been limited by their short cavity lifetime, on the order of 10 ps, compared to a thermalization time of the order of 1 ps. Our recent results \cite{Nelsen2013,Steger2013} have indicated that we can now produce structures which allow much longer lifetime, of the order of 200 ps.  Here we report on accurate measurements of this lifetime using a unique method in which we inject polariton pulses at finite momentum into a microcavity and track their motion in time and space. This allows us to observe ``slow reflection,'' in which renormalized light slows down to zero velocity, turns around, and goes back the other way.  In addition to providing a measure of the lifetime, the long-distance propagation seen here allows the possibility of beam-like polariton-interaction experiments and all-optical switching methods over long distances. 

As the technology of microcavity polaritons is now well established, much attention has turned to increasing the lifetime of the polaritons, to allow better thermalization and to allow propagation over longer distances.
The lifetime of polaritons is a function of the intrinsic photon lifetime of the cavity and the fraction of photon in the polariton states. As amply discussed elsewhere\cite{Kavokin2007}, a polariton state $|P_k \rangle$ is a superposition of an exciton state $| e_k\rangle$ and a photon state $ | \gamma_k  \rangle$,
\begin{equation}
|P_k \rangle = \alpha_k | \gamma_k  \rangle\pm \beta_k | e_k\rangle, 
\end{equation}
where $\alpha_k$ and $\beta_k$ are the $k$-dependent Hopfield coefficients. The $\pm$ signs indicate that there are two superpositions, known as the upper and lower polaritons; in the experiments reported here we focus entirely on the lower polariton branch. 
At resonance, $\alpha_k = \beta_k = 1/\sqrt{2}$, while far from resonance the polariton can be nearly fully photon-like or exciton-like. This implies that the $k$-dependent lifetime $\tau_k$ of the polaritons is given by
\begin{equation}
\frac{1}{\tau_k} =  \frac{|\alpha_k|^2}{\tau_{\rm nonrad}} + \frac{|\beta_k|^2}{\tau_{\rm cav}}.
\end{equation}
For polaritons in our GaAs-based samples, the rate of nonradiative recombination $\tau_{\rm nonrad}$ is negligible, so the lifetime is essentially entirely determined by the photon fraction and the cavity lifetime. In early polariton experiments\cite{Tassone2000, Deng2003, Balili2007, Kasprzak2006}, the cavity lifetime was on the order of 1 ps while the polariton lifetime was at most 10-15 ps, even well into excitonic detunings. This implied that polaritons would only scatter a few times on average before decaying.  In recent experiments\cite{Tosi2012,Wertz2010}, the polariton lifetime has been extended to about 30 ps. 

Our previous work \cite{Nelsen2013,Steger2013} has given estimates of the polariton lifetime in new samples of the order of 100-200 ps, allowing polaritons to propagate hundreds of microns to millimeters within the cavity and to show a sharp transition to a superfluid state.  Because of the propagation of the polaritons to long distances away from the excitation spot, the configuration of those experiments made it difficult to get an accurate measure of the lifetime. A measurement spatially restricted to the laser excitation spot would give a severe underestimate of the lifetime, because the polaritons do not stay put---they feel a force due to the cavity thickness gradient that pulls them to one side, leading them to travel hundreds of microns from the excitation spot. Therefore, to accurately measure the lifetime, a measurement must track the polaritons in space as they move. The measurements reported here do just that. These measurements confirm the earlier estimates of the lifetime but considerably reduce the uncertainty.  

\section{One-photon resonant injection}

The sample was arranged such that the gradient was aligned with the streak camera time slit, and then polaritons were injected at a large angle such that they moved directly against the gradient.  
\begin{figure}[ht]
 \begin{center}  
	\includegraphics[width=3.4in]{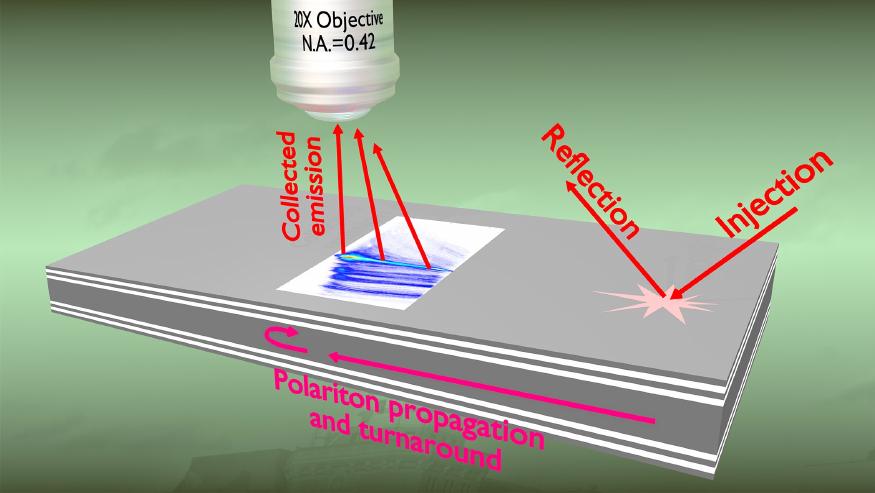}
  \caption{(color online) Diagram of experimental setup.  The sample is a microcavity polariton wafer consisting of an active GaAs/AlAs cavity layer sandwiched between distributed Bragg reflectors (DBRs) on the top and bottom.  See the supplemental information for full details of the structure.  A picosecond pulse is focused onto the sample at a large angle far outside the field of view of the collection optics.  The orientation and wavelength of the beam are selected such that resonantly created polaritons flow directly uphill against the gradient and just turn around in the field of view.  Emission from returning polaritons is not shown.  Image is not to scale.}
  \label{Fig1}
 \end{center}
\end{figure}
The experimental setup is shown in Figure~\ref{Fig1}. We used an objective with a wide field of view in addition to a large numerical aperture.  A resonantly injected picosecond pulse of polaritons was tracked as it entered the field of view, turned around and traveled away.  This occurs because the sample has a cavity thickness variation that leads to an energy gradient of the polariton.  In simple terms, one can think of the motion of the polaritons as governed by energy conservation with the following Hamiltonian, which is just the same as that of a massive object moving in a potential gradient:
\begin{equation}
H = \frac{\hbar^2k^2}{2m_{\rm eff}} - Fx.
\label{Ham}
\end{equation}
Here $m_{\rm eff}$ is the effective mass of the lower polariton branch that we observe, which depends weakly on $k$, and is equal approximately to $5\times 10^{-5}$ times the vacuum electron mass in these experiments. The force $F$ is given by the gradient in space of the $k=0$ cavity resonance energy, and is approximately equal to $10.5$ meV/mm for the section of the microcavity studied here. We will refer to ``uphill'' as moving toward higher cavity resonance energy (narrower cavity width) and ``downhill'' as moving to lower energy (wider cavity width).

This experimental setup utilizes the fact that the polaritons in these high-$Q$ samples flow over a great spatial distance and change in-plane momentum rapidly.  The lifetime of shorter-lived polaritons is more difficult to directly observe by streak camera measurements due to the overlap of any emission with the injecting laser.  Upon resonantly injecting polaritons, the created population is in the same state as the exciting laser. The initial polariton population therefore will have the same characteristics as the exciting laser and cannot be separated from it.  Observing any other state (for example by looking at cross-polarized emission) will inherently measure the scattering time of the polaritons to enter that state.  In this experiment, we rely on the fact that polaritons will flow ballistically from the point of injection to the point of detection in order to separate the observed luminescence from the reflected laser.  To the extent that this motion is ballistic, integrating the population over the observed spatial region will directly yield the population decay of the polaritons.  Unlike the case of observing luminescence from a different energy or polarization state than the initial population, this method directly follows the decay of a single population rather than relying on an average over many $k$-states.

The momentum of the injected polaritons is controlled by the angle of the laser which generates them. The angle of incidence used here was $\sim 42^\circ$, corresponding to an initial wavevector of $5.5\times 10^{-4}$ cm$^{-1}$.  After propagating uphill for over two millimeters, the polaritons enter our spatial field of view and optical collection angle.  Observing the polaritons far from the injection point reduces collection of scattered light from the laser excitation, and injecting polaritons at a large angle ensures that the reflected laser is outside the collection angle of the lens.  As the polaritons flow against the cavity gradient they lose momentum, effectively exchanging in-plane kinetic energy for confinement energy, similar to a ball rolling uphill exchanging kinetic energy for gravitational potential energy.  Since polaritons have a one-to-one relationship of in-plane momentum to external angle of emission for emitted photons, watching the luminescence change emission angle while the gas of polaritons propagates gives us a direct observation of their slowing.  Because the entire process is energy conserving, the injection laser, the ballistic polaritons and the emission all have the same wavelength.  Once the polaritons reach a turn-around point, they flow back downhill and the emission angle increases to the negative direction.

\begin{figure}[ht]
 \begin{center}  
  \includegraphics[width=3.4in]{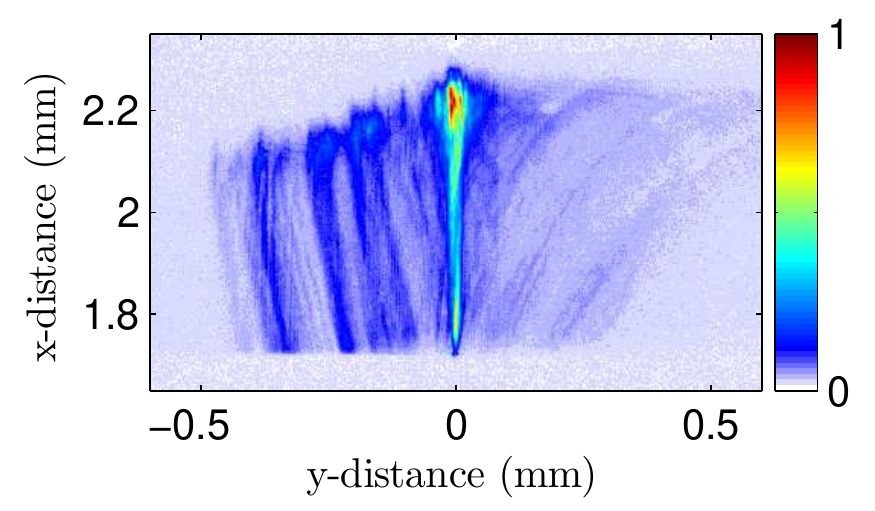}
  \caption{(color online) A time-integrated observation of the passing polariton pulse.  Coordinates are such that the point of injection is defined at ($x$,$y$)=(0,0), and the gradient is approximately toward $-x$.  Polaritons approach this field of view from the left and turn around at $x \approx2.2$ mm before flowing back to $-x$.  The sharp cutoff at $x=$1.7 mm is due to clipping in the spectrometer.}
  \label{Fig2}
 \end{center}
\end{figure}

Figure~\ref{Fig2} shows time-integrated real space emission intensity from the microcavity near the turn-around point of the polariton gas.  The coordinates in this image are such that the injection point is at roughly (0,0), and the force due to the cavity gradient is nearly directly toward $-x$.  While polaritons were injected primarily in the $+x$-direction, the initial narrow spread of momenta in the y-direction led to a spread in real space after propagation over a long distance.  At roughly $x=1.7$ mm, polaritons are seen entering the field of view, which also corresponds to the acceptance angle of the optics.  The brightest streak, directly horizontal at y=0 mm, is the trajectory of the most intense part of the injected population which was peaked at zero momentum in the $y$-direction.  Other bright streaks can be seen arcing to $\pm y$, and the entire range of states reach their respective turn-around points at  $x\leq 2.25$ mm.  The fact that there are bright streaks in this image rather than a smooth cloud suggests that the injection of the polaritons into the cavity occurs unevenly in momentum space.  The asymmetry of the cloud between $-y$ and $+y$ may be due to a slight misalignment between the cavity gradient and the injection direction.

\begin{figure}[ht]
 \begin{center}  
  \includegraphics[width=3.4in]{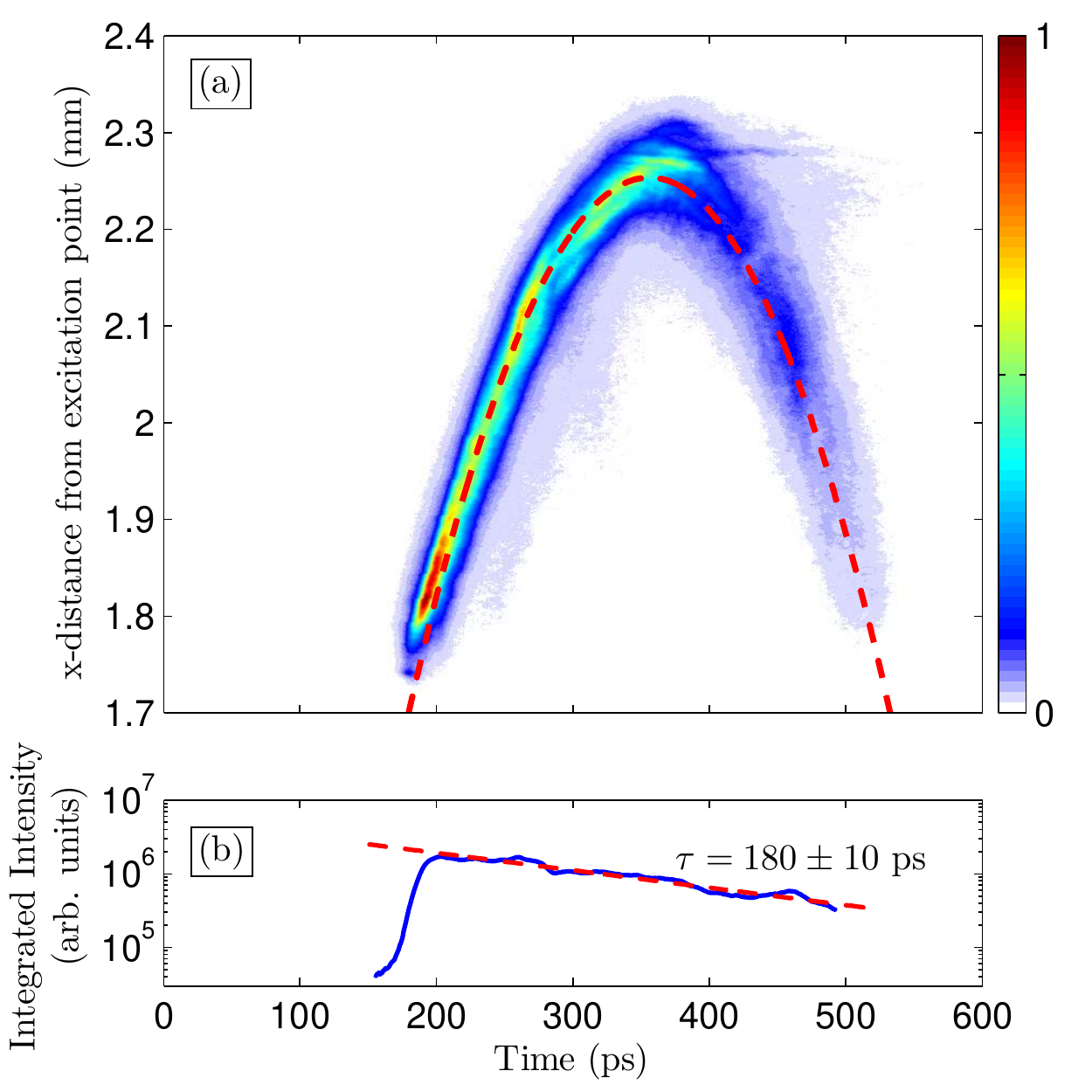}
  \caption{(color online) A time-integrated observation of the passing polariton pulse in the region of $-$150 $\mu$m $\leq y \leq$ 150 $\mu$m of fig.~\ref{Fig2}. (a) Intensity vs $x$-distance vs time of the propagating polaritons.  The dashed red line is a fit to the polariton motion as they feel a constant acceleration of $36$ mm/ns$^2$.  This acceleration is in good agreement with the expected value based on the known cavity gradient and the effective mass. (b) The polariton intensity of (a) summed in the $x$-dimension to highlight the exponential decay of the population.  The data are well fit by a single exponential decay with lifetime of $180\pm 10$ ps.}
  \label{Fig3}
 \end{center}
\end{figure}

To measure the lifetime, the bright jet of polaritons was time-resolved using a Hamamatsu streak camera.  To facilitate this, the sample was initially installed such that the gradient was aligned with the horizontal time slit on the streak camera.  This enabled us to track a single jet of polaritons while they propagate against the gradient, turn around, and travel backwards, as shown in Figure~\ref{Fig3}(a).  The vertical distance axis in this figure corresponds to the horizontal $x$-axis in Figure~\ref{Fig2}.  The trajectory of the polaritons is easily seen in the data, which in this region is well described by a parabolic fit, as expected for the Hamiltonian (\ref{Ham}), which is equivalent to that of a ball moving with a constant force due to gravity.  Indeed, these data directly demonstrate the in-plane velocity and acceleration of the polaritons during their trajectory.  One should note that this region of observation is already more than a millimeter and nearly 200 ps from the injection point, indicating that these polaritons are propagating farther and persisting longer than those in earlier samples, even without confinement in 1D structures, such as used in Ref.~\cite{Wertz2010}.

A simple analysis of this data yields the polariton lifetime after summing in the spatial dimension, as shown in Figure~\ref{Fig3}(b).  The data are well fit by a single exponential with a lifetime of $180 \pm 10$ ps.  For the region of the sample observed in Figure~\ref{Fig3}, the detuning of the polariton corresponds to the lower polariton approximately 75\% photonic.  (Although the polaritons move long distances, their detuning does not change much because they stay at the same energy.) From this we estimate that the cavity photon lifetime is approximately 135 ps, which corresponds to a $Q$-factor of over 320,000.

It should be noted that this lifetime measurement may still be an underestimation of the lifetime.  Close inspection of Figure~\ref{Fig2} reveals that individual jets of polaritons are still spreading out from the central jet.  A population with some spread in initial momenta perpendicular to the cavity gradient must spread out horizontally while propagating uphill.  The fraction of polaritons that move out of our field of view will lead to an underestimation of the lifetime.  This error can be compensated for by using a narrower time slit to cut out adjacent jets at early times; however, narrow slit widths can result in errors that will either underestimate or overestimate the polariton lifetime if the entirety of the main jet is not aligned with the time slit.  In this experiment, data was collected over a range of slit widths from 50 to 300 $\mu$m with consistent results.

\section{Two-photon resonant injection}

\begin{figure*}[ht]
 \begin{center}  
  \includegraphics[width=6in]{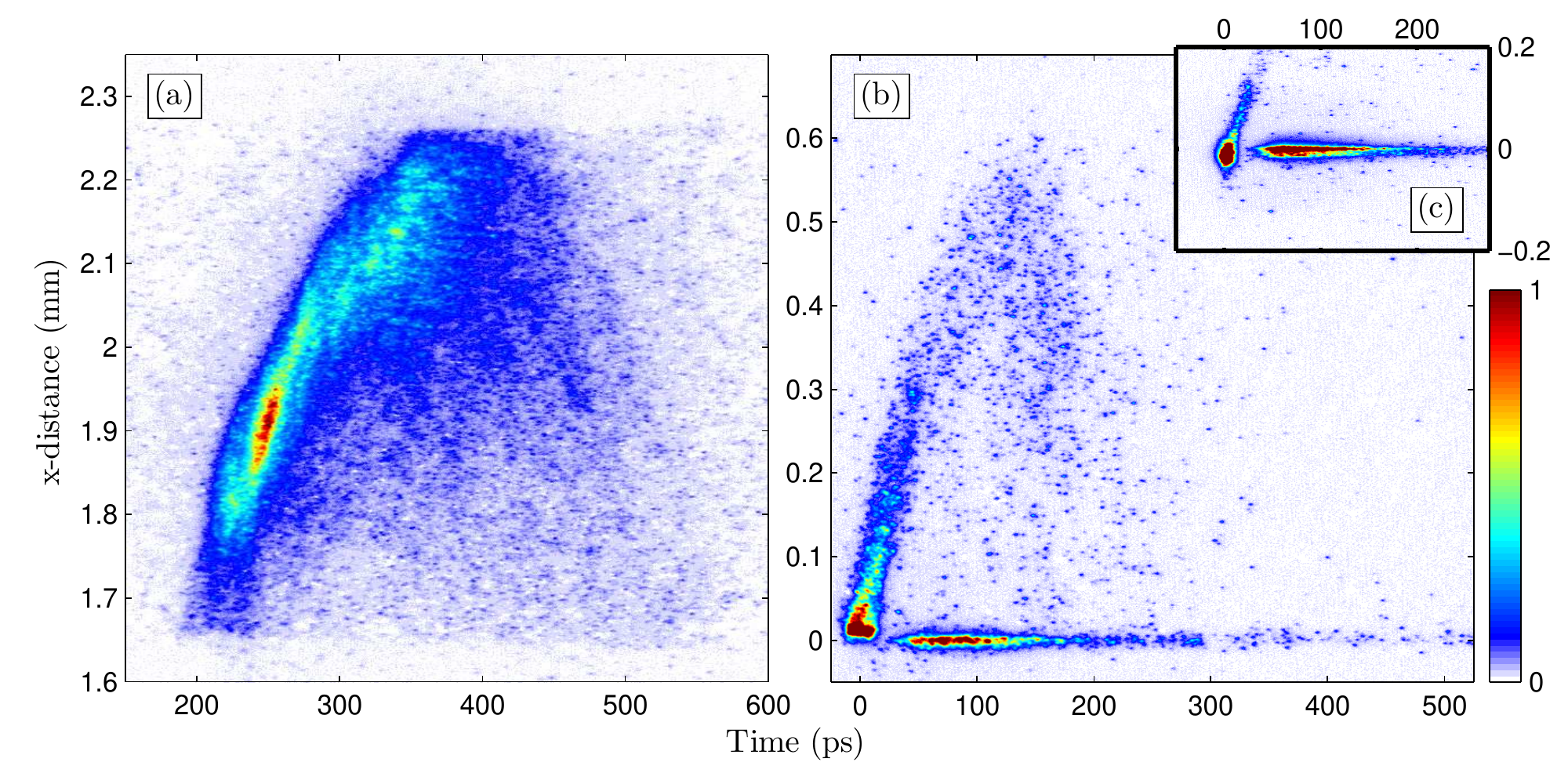}
  \caption{(color online) Time-resolved polariton emission following two-photon excitation.  In these data, a 200-femtosecond pulse tuned to twice the wavelength of the lower polariton is used to generate a population of polaritons.  (a) Time-resolved image for polaritons excited at a large angle far away from the observation point just as in Figure~\ref{Fig3}.  Unlike the previous data where polaritons are created in a narrow bandwidth related to that of the resonantly exciting picosecond laser, this population exists in a spread of spectral states and thus follows a range of trajectories in x vs t. (b)  Time-resolved image for polaritons injected by the femtosecond laser at normal incidence at the point of observation, directly through the microscope objective.  The full NA of the objective is used, meaning that we can expect to see polaritons filling a range of momentum states according to the two-photon phase matching condition of the absorption.  The long lived stationary population is substrate luminescence due to the pump penetrating through the microcavity structure. It should be noted that the point of injection is partially clipped in this figure to allow for imaging the entire turn-around.  The inset (c) shows the same conditions as (b) except with the pump spot centered and unclipped.}
  \label{Fig4}
 \end{center}
\end{figure*}

In another set of experiments, we generated polaritons via a 200-fs pump pulse tuned to one-half the energy of the polaritons.  Such experiments will be presented in more depth elsewhere, but we can see here that the propagation of long-lived polaritons can be used to probe the dynamics of two-photon generated polaritons.  We conducted two experiments. First, we recreated geometry of the one-photon experiments discussed above by injecting polaritons uphill and observing the turn-around point, as shown in Figure~\ref{Fig4}(a). Second, we injected polaritons at normal incidence, directly at the point of observation, as shown in Figure~\ref{Fig4}(b).

Figure~\ref{Fig4}(a) clearly shows characteristics similar to the one-photon resonant injection data presented in Figure~\ref{Fig3}, with polaritons entering the field of view at $\sim$ 200 ps after the injecting laser pulse and following a roughly parabolic trajectory.  However, the trajectory is very broad with a poorly defined turn-around point.  This can be understood by recognizing that the initial population is created in a range of energy states due to the spectral width of the exciting femtosecond pulse.  Since there is a broad range of photon energy and momenta in the pump, a range of polariton states can satisfy energy and momentum conservation in the two photon absorption.   This method of observing polariton propagation can therefore enable us to characterize the initial polariton population.

In Fig.~\ref{Fig4}(b) we present data from a different setup employing two-photon generation of polaritons.  In this case, we pump at normal incidence directly through the microscope objective used to image the luminescence. Since the pump laser has wavelength far from the polariton wavelength, there is no difficulty with scattered laser light.  Since the entire NA of the objective was used, the angle of incidence of the pump light ranged over $\pm$ 20$^\circ$ even though the intensity was maximum at 0$^\circ$ and the laser was spectrally tuned to the $k=0$ state.  The data indicate that polaritons created directly via two-photon absorption are peaked at an initial wavevector uphill, with no $k=0$ polaritons created initially.  The long-lived population stationary at $x=0$ is substrate luminescence excited by two-photon absorption.  Figure~\ref{Fig4}(c) shows the same data as Figure~\ref{Fig4}(b), except that the image is centered to eliminate the clipping at $x<0$ in (b). The majority of the injected polaritons have finite $k$, even though the pump light was centered at $k=0$.  This supports the view that that two-photon absorption of the polaritons, which should be forbidden at $k=0$ due to the selection rules, becomes allowed at finite $k$, due to valence-band mixing with the higher-lying light-hole states. 

These results show clearly that two-photon resonant generation of polaritons is possible.  One can expect very strong nonlinear effects from microcavity polaritons due to the strong interaction of the cavity mode with the quantum well exciton\cite{Pellegrini1999,Lei2011}.  We point out that there is no comparable one-photon excitation experiment---such an experiment will not work because the exciting laser will be reflected directly back into the imaging system.

\section{Conclusions}

Polaritons can be viewed as ``renormalized photons,'' especially in the region of the cavity where the the detuning makes the polaritons mostly photon-like. As mentioned above, the behavior we have seen here can thus be viewed as a type of ``slow light,'' or ``slow reflection,'' in which the photons decelerate from $\sim$ 3.5\% of the speed of light to a full stop and then go back the other way. This behavior is expected for light in a wedge-shaped cavity, without any need for the excitonic part of the polaritons. However, it has been hard to directly observe, because one must have very high $Q$ and fast time resolution to track the motion of the photons. These measurements show that the photons can truly be viewed as having effective mass and feeling a force. 

With such long distance propagation and long lifetime, it is now possible to construct experiments in which two or more beams are used and caused to interact. This could be used to directly measure the polariton-polariton interactions and also for schemes of optical gating using polaritons, as presented e.g. in Ref.~\cite{Ballarini2013}.

Additionally, we have shown that two-photon injection of polaritons is a very relevant phenomena in GaAs microcavities, but there is a noticeable dependence of the absorption strength on the angle of incidence.  By temporally- and spatially-resolving the propagation of polaritons after generation we can probe the mechanisms by which different states are occupied by this nonlinear absorption.

\section*{Funding Information}

The work at the University of Pittsburgh was supported by the National Science Foundation under grants PHY-1205762 and ECCS-1243778.  The work at Princeton University was partially funded by the Gordon and Betty Moore Foundation as well as the National Science Foundation MRSEC Program through the Princeton Center for Complex Materials (DMR-0819860).


\begin{thebibliography}{10}
\newcommand{\enquote}[1]{``#1''}

\bibitem{Weisbuch1992}
C.~Weisbuch, M.~Nishioka, A.~Ishikawa, and Y.~Arakawa, \enquote{{Observation of
  the coupled exciton-photon mode splitting in a semiconductor quantum
  microcavity},} Physical Review Letters \textbf{69}, 3314--3317 (1992).

\bibitem{Kasprzak2006}
J.~Kasprzak, M.~Richard, S.~Kundermann, A.~Baas, P.~Jeambrun, J.~M.~J. Keeling,
  F.~M. Marchetti, M.~H. Szymańska, R.~Andr\'{e}, J.~L. Staehli, V.~Savona,
  P.~B. Littlewood, B.~Deveaud, and L.~S. Dang, \enquote{{Bose–Einstein
  condensation of exciton polaritons},} Nature \textbf{443}, 409--14 (2006).

\bibitem{Balili2007}
R.~Balili, V.~Hartwell, D.~Snoke, L.~Pfeiffer, and K.~West,
  \enquote{{Bose-einstein condensation of microcavity polaritons in a trap},}
  Science \textbf{316}, 1007--1010 (2007).

\bibitem{Lagoudakis2011}
K.~G. Lagoudakis, F.~Manni, B.~Pietka, M.~Wouters, T.~C.~H. Liew, V.~Savona,
  A.~V. Kavokin, R.~Andr\'{e}, and B.~Deveaud-Pl\'{e}dran, \enquote{{Probing
  the Dynamics of Spontaneous Quantum Vortices in Polariton Superfluids},}
  Physical Review Letters \textbf{106}, 115301 (2011).

\bibitem{Amo2011}
A.~Amo, S.~Pigeon, D.~Sanvitto, V.~G. Sala, R.~Hivet, I.~Carusotto,
  F.~Pisanello, G.~Lemenager, R.~Houdre, E.~Giacobino, C.~Ciuti, and
  A.~Bramati, \enquote{{Polariton Superfluids Reveal Quantum Hydrodynamic
  Solitons},} Science \textbf{332}, 1167--1170 (2011).

\bibitem{Nelsen2013}
B.~Nelsen, G.~Liu, M.~Steger, D.~W. Snoke, R.~Balili, K.~West, and L.~Pfeiffer,
  \enquote{{Dissipationless Flow and Sharp Threshold of a Polariton Condensate
  with Long Lifetime},} Physical Review X \textbf{3}, 041015 (2013).

\bibitem{Steger2013}
M.~Steger, G.~Liu, B.~Nelsen, C.~Gautham, D.~W. Snoke, R.~Balili, L.~Pfeiffer,
  and K.~West, \enquote{{Long-range ballistic motion and coherent flow of
  long-lifetime polaritons},} Physical Review B \textbf{88}, 235314 (2013).

\bibitem{Kavokin2007}
A.~V. Kavokin, J.~J. Baumberg, G.~Malpuech, and F.~P. Laussy,
  \emph{{Microcavities}} (Oxford University Press, New York, 2007).

\bibitem{Tassone2000}
F.~Tassone and Y.~Yamamoto, \enquote{{Lasing and squeezing of composite bosons
  in a semiconductor microcavity},} Physical Review A \textbf{62}, 063809
  (2000).

\bibitem{Deng2003}
H.~Deng, G.~Weihs, D.~Snoke, J.~Bloch, and Y.~Yamamoto, \enquote{{Polariton
  lasing vs. photon lasing in a semiconductor microcavity},} Proceedings of the
  National Academy of Sciences of the United States of America \textbf{100},
  15318--15323 (2003).

\bibitem{Tosi2012}
G.~Tosi, G.~Christmann, N.~G. Berloff, P.~Tsotsis, T.~Gao, Z.~Hatzopoulos,
  P.~G. Savvidis, and J.~J. Baumberg, \enquote{{Sculpting oscillators with
  light within a nonlinear quantum fluid},} Nature Physics \textbf{8}, 190--194
  (2012).

\bibitem{Wertz2010}
E.~Wertz, L.~Ferrier, D.~Solnyshkov, R.~Johne, D.~Sanvitto, A.~Lema\^{\i}tre,
  I.~Sagnes, R.~Grousson, A.~V. Kavokin, P.~Senellart, G.~Malpuech, and
  J.~Bloch, \enquote{{Spontaneous formation and optical manipulation of
  extended polariton condensates},} Nature Physics \textbf{6}, 860--864 (2010).

\bibitem{Pellegrini1999}
V.~Pellegrini, R.~Colombelli, I.~Carusotto, F.~Beltram, S.~Rubini, R.~Lantier,
  A.~Franciosi, C.~Vinegoni, and L.~Pavesi, \enquote{{Resonant second harmonic
  generation in ZnSe bulk microcavity},} Applied Physics Letters \textbf{74},
  1945--1947 (1999).

\bibitem{Lei2011}
S.~Lei, Y.~Yao, Z.~Li, T.~Yu, and Z.~Zou, \enquote{{Design and theoretical
  analysis of resonant cavity for second-harmonic generation with high
  efficiency},} Applied Physics Letters \textbf{98}, 1--3 (2011).

\bibitem{Ballarini2013}
D.~Ballarini, M.~{De Giorgi}, E.~Cancellieri, R.~Houdr\'{e}, E.~Giacobino,
  R.~Cingolani, A.~Bramati, G.~Gigli, and D.~Sanvitto, \enquote{{All-optical
  polariton transistor.}} Nature communications \textbf{4}, 1778 (2013).

\end{thebibliography}



\newpage



%
%




\begin{center}
    \textbf{\LARGE SUPPLEMENTAL INFORMATION}
 \end{center}

\section*{Sample Details}
This sample is the same as used in Refs \cite{Nelsen2013,Steger2013}.  A $3\lambda/2$  microcavity contains three sets of four GaAs quantum wells located at the antinodes of the cavity mode.  The QWs are 70 \AA\ pure GaAs embedded in pure AlAs barriers at least 30 \AA\ thick. The optical mode is confined between distributed Bragg reflectors made of AlAs/Al$_{0.2}$Ga$_{0.8}$As with 32 pairs on the top surface and 40 pairs on the bottom surface.  Molecular beam epitaxial growth of the sample leads to an inherent wedge to the cavity thickness, resulting in a gradient of the cavity as well as exciton energies.  The polariton exhibits a Rabi coupling of 6 meV at 5 K; the cavity mode gradient is 13 meV/mm and the exciton gradient is 1.5 meV/mm.

\section*{Methods}

The sample was held in a cold-finger cryostat at 5 K for all experiments.

Emission was collected using a N.A.=0.42 microscope objective.  A preliminary imaging lens permitted spatial filtering of the real space image data, and a subsequent iris in the Fourier image plane permitted filtering of the emission angle.  Secondary lenses could be exchanged to image either the real-space or angle-resolved emission.  Luminescence was imaged through a spectrometer onto either a standard CCD or onto a Hamamatsu streak camera.

\section*{One-photon injection}

In the resonant injection experiment, polaritons were resonantly injected at $\lambda = 778$ nm with a picosecond laser far on the photonic side of the sample.  The injected state had a detuning of approximately -2.6 meV and corresponded to an external angle of roughly 42$^{\circ}$ ($k_{\|}=5.5\times 10^{-4}$ cm$^{-1}$).  The sample and pump laser were arranged such that the polaritons were moving anti-parallel to the cavity gradient, which was aligned with the time slit.  The angle of incidence was larger than the collection angle of the optics, so the reflected beam was not collected.  Additionally, the pump spot was spatially outside the field of view such that scattered light was not collected.  At a distance of approximately 2 mm from the injection point, emission entered the collection range of the optics.  At the turn-around point, the polaritons are more photonic with a detuning of -7.4 meV which corresponds to a photon fraction of 75\%.

\section*{Two-photon injection}

For the two-photon injection experiment, a 200-fs pulse generated by a Coherent OPA system was tuned to one-half the energy of the desired polariton transition.  For the 40$^\circ$ injection case, the laser was tuned to excite at $\lambda = 778$ nm.  For the 0$^\circ$ injection case, the laser was tuned to excite at $\lambda = 780$ nm.

\section*{Analysis}

The x-distance from the injection point for the 40$^\circ$ injection cases was estimated as follows: the $x$-distance in figs.~\ref{Fig2},\ref{Fig3}(a), and \ref{Fig4}(a) was determined from the fit of the polariton $x$ vs $t$ trajectory presented in fig.~\ref{Fig3}(a).  Extrapolation of this fit back to time $t=0$ as determined by locating scattered laser light determines the initial position of excitation.  This initial excitation position is consistent with the sample parameters and injection conditions.  This method assumes that the acceleration of the polaritons is strictly constant from creation to turn around.  Variation in the acceleration due to a non-constant energy gradient in addition to the changing mass of the polariton implies uncertainty on the overall offset of this axis, but the spatial magnification was measured directly.

\end{document}